\documentclass[12pt]{article}
\usepackage{cite} 
\usepackage{epsf} 
\usepackage{epsfig} 
\usepackage{amsfonts,latexsym}
\usepackage{epsfig,multicol}
\begin{document}
\newcommand{\de}{\delta}\newcommand{\ga}{\gamma}
\newcommand{\e}{\epsilon} \newcommand{\ot}{\otimes}
\newcommand{\be}{\begin{equation}} \newcommand{\ee}{\end{equation}}
\newcommand{\ba}{\begin{array}} \newcommand{\ea}{\end{array}}
\newcommand{\beq}{\begin{equation}}\newcommand{\eeq}{\end{equation}}
\newcommand{\tmod}{{\cal T}}\newcommand{\amod}{{\cal A}}
\newcommand{\bemod}{{\cal B}}\newcommand{\cmod}{{\cal C}}
\newcommand{\dmod}{{\cal D}}\newcommand{\hmod}{{\cal H}}
\newcommand{\s}{\scriptstyle}\newcommand{\tr}{{\rm tr}}
\newcommand{\einsop}{{\bf 1}}
\def\R{\overline{R}} \def\doa{\downarrow}
\def\dag{\dagger}
\def\ve{\epsilon}
\def\si{\sigma}
\def\ga{\gamma}
\def\no{\nonumber}
\def\le{\langle}
\def\re{\rangle}
\def\lt{\left}
\def\rt{\right}
\def\dwn{\downarrow}   
\def\up{\uparrow}
\def\dag{\dagger}
\def\nonum{\nonumber}
\newcommand{\reff}[1]{eq.~(\ref{#1})}

\title{Magnetic susceptibility of an integrable anisotropic spin ladder system}

\author{A.~P.~Tonel$^{1}$,
S.~R.~Dahmen$^{1}$, A. Foerster$^{1}$ and A. L. Malvezzi$^{2}$
\vspace{1.0cm}\\
$^{1}$Instituto de F\'{\i}sica da UFRGS \\
Av. Bento Gon\c{c}alves 9500, Porto Alegre, RS - Brazil
\vspace{0.5cm}
\vspace{0.5cm}\\
$^{2}$Departamento de F\'{\i}sica da UNESP \\
Av. Eng. Luiz Edmundo Carrijo Coube, s/n,  Bauru, SP-Brazil
}
\maketitle

\begin{abstract}
 We investigate the thermodynamics of an integrable spin ladder model
 which possesses a  free parameter besides the rung and leg
 couplings. This model is exactly solvable by means of the Bethe Ansatz
 and exhibits a phase transition between a gapped and a gapless
 spin excitation spectrum. The magnetic susceptibility
 is obtained numerically and its dependence on the anisotropy
 parameter is determined. A comparison between the spin gap
 obtained from the susceptibility curve and the one obtained from the
 Bethe Ansatz Equations is made and a good agreement found.  
 A connection with the compounds $KCuCl_3$ \cite{r1,r2},
 $Cu_2(C_5H_{12}N_2)_2Cl_4$ \cite{r3,r4,r5} and $(C_5H_{12}N)_2CuBr_4$
 \cite{r6} in the strong coupling regime is made and our
 results for the magnetic susceptibility fit the experimental
 data remarkably well.
\end{abstract}


\vfil\eject
\section{Introduction}

The study of spin ladders, both from an experimental as well as from
a theoretical point of view  has gained much attention in the last
few years and has by now a literature of its own.
On the one hand with 
the discovery of high--Tc superconductivity in doped
cuprate  materials \cite{r7}, a tremendous
effort has been made to understand the physics of these 
compounds, in which the $Cu$--$O$ ions build up a
two-dimensional lattice. This task is made more difficult
by the lack of exactly integrable models as compared to 1--D
where an abundance of such models
has been known for decades. On the other hand spin ladders
are prototypes of quasi one-dimensional systems and as such represent an
excellent proving ground for studying the transition from
the more amenable and better understood physics of one--dimensional
quantum systems to two--dimensional behavior. They
exhibit
{\it antiferromagnetic} (AF) ground states, where at each site of
the lattice (in the undoped state) an
electron occupies one of two spin states \cite{r8}. As such they 
can be reasonably well approximated
by the archetypical AF Heisenberg model
or some suitable generalization thereof.
In one dimension
the Heisenberg model is exactly solvable via the Bethe Ansatz and
from the solution it is known
that in the AF regime, up to a given threshold of the spin exchange interaction,
the elementary spin excitations are {\it gapless} (beyond the threshold
one has a N\'eel-like groundstate with gapped excitations) \cite{bethe,dagotto}.
The existence of a spin gap is critical for the occurence of
superconductivity under doping, whereby the holes introduced through doping
undergo a Bose-Einstein condensation.
By introducing the concept of a
ladder model this apparent contradiction is resolved, since these systems
allow for the formation of singlet states along the rungs which in turn are
responsible for the appearance of a spin gap. (Strictly speaking
the singlet states can only form when there is an even number of 
legs \cite{rice,white}). Recently, with the rapid progress
being  made in nano-engineering, many
different species of ladder compounds such as $S_rCu_2O_3$, $
La_{1-x}Sr_xCuO_{2.5}$, $Sr_{14-x}Ca_xCu_{24}O_{41}$  and other
organic compounds \cite{r2} have been investigated in this context.
This theory is now supported by a substantial body of experimental evidence.

However, in contradistinction to its one-dimensional counterpart,
the two--dimensional Heisenberg ladder model cannot be solved exactly.
So in the search
for exactly integrable models which might give one a better insight
into the physics of ladder systems, many authors 
have considered generalized models which incorporate additional
interaction terms without violating integrability
\cite{r9,r10,r10a,r10b,r10c,r10d,r10e,r10f,r10g,r11,r12,r13}.
Notwithstanding the fact that these
models exhibit realistic physical properties such as 
the existence of a spin gap \cite{r13,r14} 
and the magnetization plateaus at fractional
values of the total magnetization \cite{r15}, up to our knowledge none
has been used to predict physical quantities that
could be compared directly with experimental data.

The main goal of the present work is to show that integrable ladder models
can be used to model experimental data. In particular, we discuss
a generalized integrable spin ladder with incorporates an additional anisotropy
parameter besides the rung and leg interactions and show that the magnetic
susceptibility as obtained from this model as a function of temperature
fits well the experimental data
for the ladder compounds $KCuCl_3$ \cite{r1,r2}, $Cu_2(C_5H_{12}N_2)_2Cl_4$
\cite{r3,r4,r5} and  $(C_5H_{12}N)_2CuBr_4$ \cite{r6}.
The model is exactly solvable by the Bethe Ansatz
and  reduces to the model introduced by Wang \cite{r13,r14}
for some special limit of this extra parameter.
This additional free parameter, which we call $t$, brings about an
anisotropy--dependent critical line through an explicit functional dependence
of the excitation gap.
The thermodynamics of the model is investigated and the magnetic 
susceptibility curve as a function of the temperature is obtained.
The influence of this anisotropy parameter on the thermodynamics of the
model is also determined. A comparison between the spin gap obtained from
the susceptibility curve and that one obtained from the Bethe ansatz
equations is performed and a good agreement is found. 
 
The paper is divided as follows: In
section $2$ we describe the model and its Bethe Ansatz solution. Section
$3$ discusses the thermodynamical properties of the model and is also
devoted to a comparison between the experimental results and our findings.
We conclude the paper with a brief summary.
\section{The model}
Let us begin by presenting the integrable anisotropic  spin ladder model in 
the presence of an external magnetic field $h$\cite{r12},
whose Hamiltonian reads
\be
H =\sum_{j=1}^{L} \biggl[J_l \, h_{j,j+1} +
{\frac{J_r}{2}}
\left( \vec{\sigma_{j}}.\vec{\tau_{j}}-1 \right) 
-\frac{h}{2}\left( \sigma_j^z + \tau_j^z \right)\,
\biggr]
\label{ha}
\ee
where
\begin{eqnarray}
&h_{j,j+1}&=\frac{1}{4}(1+\sigma_{j}^{z}\sigma_{j+1}^{z})(1+\tau_{j}^{z}\tau_{j+1}^{z}) \,
+ 
(\sigma_{j}^{+}\sigma_{j+1}^{-}+\sigma_{j}^{-}\sigma_{j+1}^{+})(\tau_{j}^{+}\tau_{j+1}^{-}+\tau_{j}^{-}\tau_{j+1}^{+})\quad \nonumber \\ & & +
 \frac{1}{2}(1+\sigma_{j}^{z}\sigma_{j+1}^{z})(t^{-1}\tau_{j}^{+}\tau_{j+1}^{-}+
t\tau_{j}^{-}\tau_{j+1}^{+})+
 \frac{1}{2}(t^{-1}\sigma_{j}^{+}\sigma_{j+1}^{-}+
t\sigma_{j}^{-}\sigma_{j+1}^{+})(1+\tau_{j}^{z}\tau_{j+1}^{z}) \; .
\nonumber 
\end{eqnarray}
Periodic boundary conditions are imposed along the legs.
As usual $\vec{\sigma_{j}}$ and $\vec{\tau_{j}}$ are Pauli matrices acting
on site $j$ of the upper and lower legs respectively, $J_r(J_l)$ is the
strength of the rung coupling (leg coupling) and
$t$ is a free parameter which introduces an
anisotropy in the leg and in the interchain interactions. 
Throughout this paper  $L$ stands for the number of rungs
(the length of the ladder).
By setting $t \rightarrow 1$ in Eq. (\ref{ha}) one recovers the
isotropic model of Wang \cite{r14} based on the $SU(4)$ symmetry
(strictly speaking Wang's model is $SU(4)$--invariant only in
the absence of the rung interactions).
The Hamiltonian is invariant under the interchange of the legs, {\it i.e.} 
$\vec{\sigma_{j}}\leftrightarrow \vec{\tau_{j}}$. 
Moreover, under spin inversion on both leg spaces it is also
invariant under the exchange $t\leftrightarrow t^{-1}$. 
The energy eigenvalues of (\ref{ha}) are given by 
\begin{equation}
\label{energy}
\frac{E}{J_l}=-\sum_{j=1}^{M_{1}}\biggl(\frac{1}{\lambda_{j}^{2}+1/4}-2\;
\frac{J_r}{J_l} +\frac{h}{J_l}\biggr)+\left(1-2\;\frac{J_r}{J_l} \right)L +\frac{h}{J_l}(M_2+M_3)
\end{equation}
where the $\lambda_{j}$'s are solutions of the three--level nested 
Bethe ansatz equations
\begin{eqnarray}
t^{(L-2M_{3})}\left(\frac{\lambda_{j}-i/
2}
{\lambda_{j}+i/2}\right)^{L}&=&\prod_{l\neq
j}^{M_{1}}\frac{\lambda_{j}-\lambda_{l}-i}
{\lambda_{j}-\lambda_{l}+i}
\prod_{\alpha=1}^{M_{2}}\frac{\lambda_{j}-\mu_{\alpha}+i/2}
{\lambda_{j}-\mu_{\alpha}-i/2} \nonumber \\
t^{(L-2M_{3})}\prod_{\beta\neq 
\alpha}^{M_{2}}\frac{\mu_{\alpha}-\mu_{\beta}-i}
{\mu_{\alpha}-\mu_{\beta}+i}&=&\prod_{j=1}^{M_{1}}\frac{\mu_{\alpha}-\lambda_{j}-i/2}
{\mu_{\alpha}-\lambda_{j}+i/2}
\prod_{\delta =1}^{M_{3}}\frac{\mu_{\alpha}-\nu_{\delta}-i/2}
{\mu_{\alpha}-\nu_{\delta}+i/2}
\label{bae} \\
t^{(L+2M_{2}-2M_{1})}\prod_{\gamma \neq
\delta}^{M_{3}}
\frac{\nu_{\delta}-\nu_{\gamma}-i}{\nu_{\delta}-\nu_{\gamma}+i} &=&
\prod_{\alpha =1}^{M_{2}}\frac{\nu_{\delta}-\mu_{\alpha}-i/2}
{\nu_{\delta}-\mu_{\alpha}+i/2}
\nonumber
\end{eqnarray}
For the sake of convenience we take from here on $J_l=1$.
From an analysis of these equations it follows that for 
$J_r > 1+\frac{1}{2}(t+\frac{1}{t})$ the reference state becomes the 
ground state and any excitation is gapped. In this region the 
ground state is given by a product of rung singlets, indicating
a dimerization along the rungs.The energy gap $\Delta$
can be calculated using the exact  Bethe ansatz solution and has the form
\begin{equation}
\Delta=2\biggl(J_r -1-\frac{h}{2}-\frac{1}{2}(t+\frac{1}{t})\biggr)  .
\label{gap}
\end{equation}
By solving $\Delta = 0$ for $J_r$ we find the critical value
${J_r}^c=1+\frac{1}{2}(t+\frac{1}{t})$, indicating the critical 
line at which the quantum phase transition from the 
dimerized phase to the gapless phase occurs.
This result is depicted in Fig. 1. 

Notice that by choosing
$t$ real one has ${J_r}^c \geq 2$ which corresponds to the
strong coupling limit. We therefore expect our anisotropic model
to be a good candidate for describing some
organic ladder compounds, such as $KCuCl_3$, $Cu_2(C_5H_{12}N_2)_2Cl_4$,
$TlCuCl_3$ and $(C_5H_{12}N)_2CuBr_4$.
\begin{figure}[ht]
\begin{center}
\epsfig{file=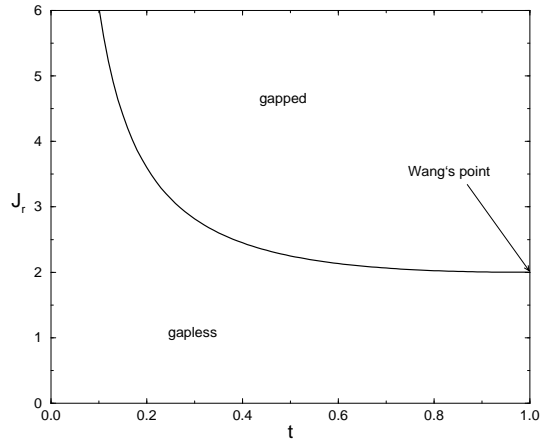,width=6cm,height=7cm,angle=-90}
\caption{ The phase diagram of (\ref{ha}) in the ($J_r$ $\times$ $t$)--plane
in units of $J_l$. The arrow indicates
Wang's point (t=1). The curve ${J_r}^c = 1+(t+1/t)/2$
divides the gaped from the gapless phase.}
\end{center}
\end{figure}
\section{Comparison with experimental data} 

Our main goal is to calculate the magnetic susceptibility $\chi$ of our
exactly integrable ladder Hamiltonian and compare it to experimental
data on the compounds mentioned before.

In order to do this we block--diagonalize the Hamiltonian Eq. (\ref{ha})
for a sequence of finite chains (up to $6$ rungs or $12$ sites) and for
different values of the coupling $J_r$ and 
anisotropy $t$. With these spectra we generate a sequence of
finite--size susceptibilities $\chi^{(\mbox{\tiny L})}$ 
as a function of the temperature $T$ and
then apply a standard Bulirsch-Stoer
acceleration technique \cite{r16} to the sequence 
in order to determine the susceptibility in the thermodynamic limit
$L\rightarrow\infty$. An example for $J_r = 7$ and $t=1$ is presented
in Fig. 2.
%
%
\begin{figure}[ht]
\begin{center}
\epsfig{file=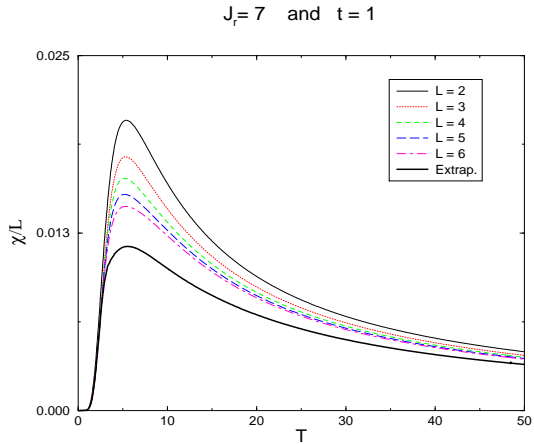,width=6cm,height=7cm,angle=-90}
\caption{The finite-size susceptibility per site for the case
$J_r = 7$ and $t=1$ for varying number of rungs $L$ and the
extrapolated curve.}
\end{center}
\end{figure}
%
%
In order to compare our results with
experimental data available on a series of compounds, a few
remarks on the behavior of the susceptibility as a function of $J_r$ and $t$
are necessary, as well as the procedure used for choosing the
values of our parameters. First we observe that for a fixed value of $t$
there is a 'smoothing out' of the susceptibility curve for increasing $J_r$
while for fixed $J_r$  the magnetic susceptibility increases
with decreasing $t$, as can be seen in Fig. $3a$. 
%
%
\begin{figure}
\begin{center}
\begin{tabular}{cc}
            &             \\
    (a)& (b)    \\
\includegraphics[scale=0.3,angle=-90]{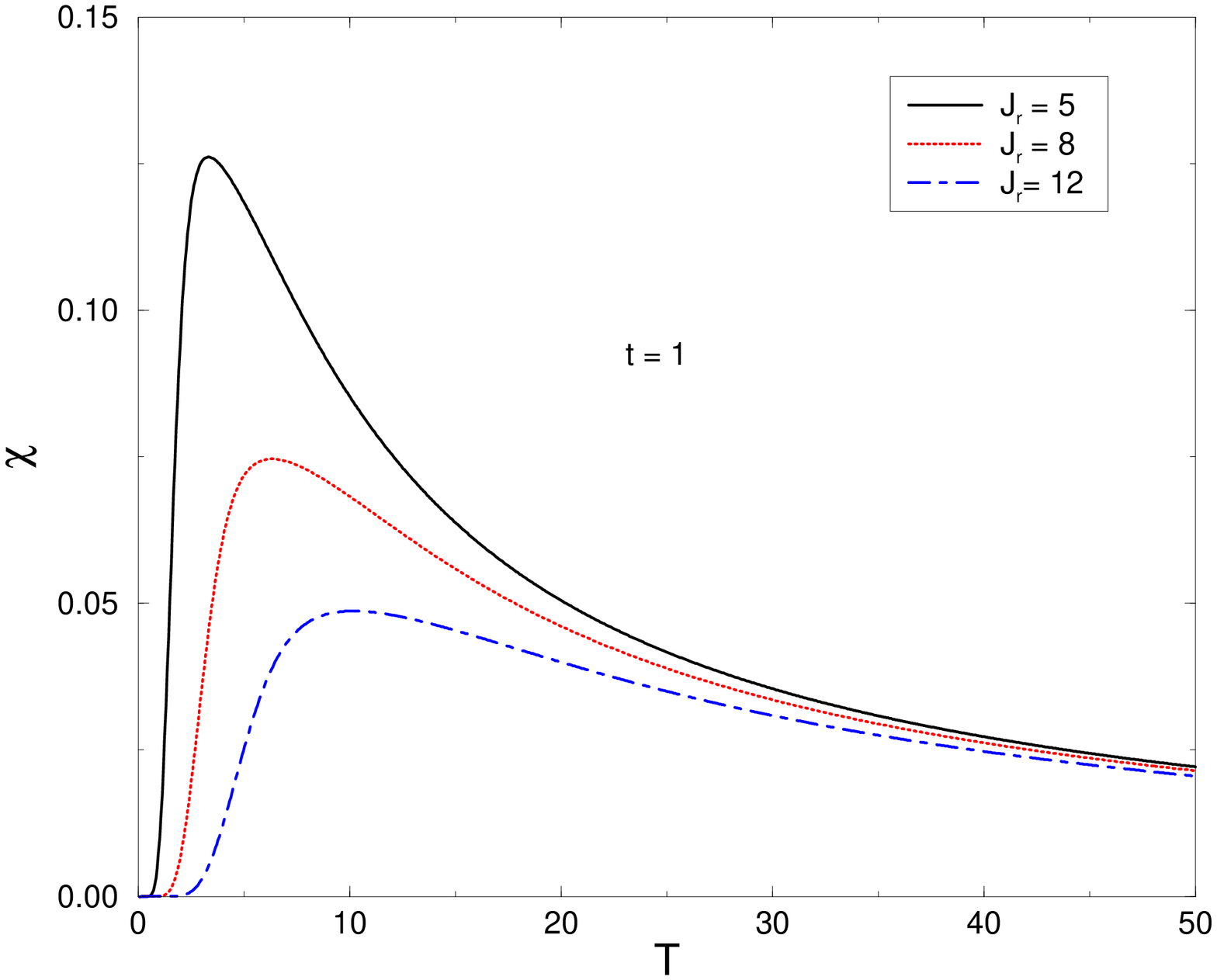}&            
\includegraphics[scale=0.3,angle=-90]{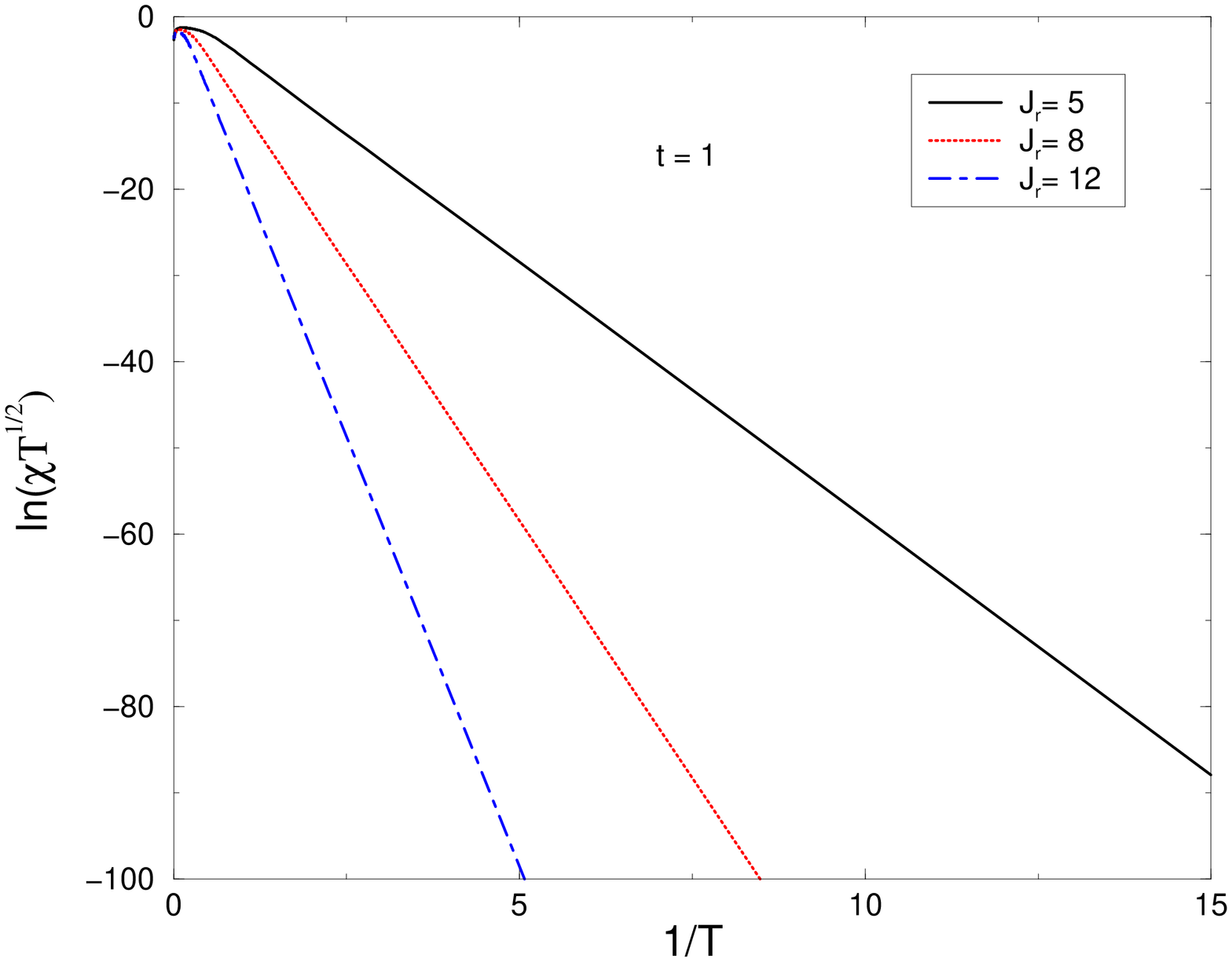} \\
           &              \\
\includegraphics[scale=0.3,angle=-90]{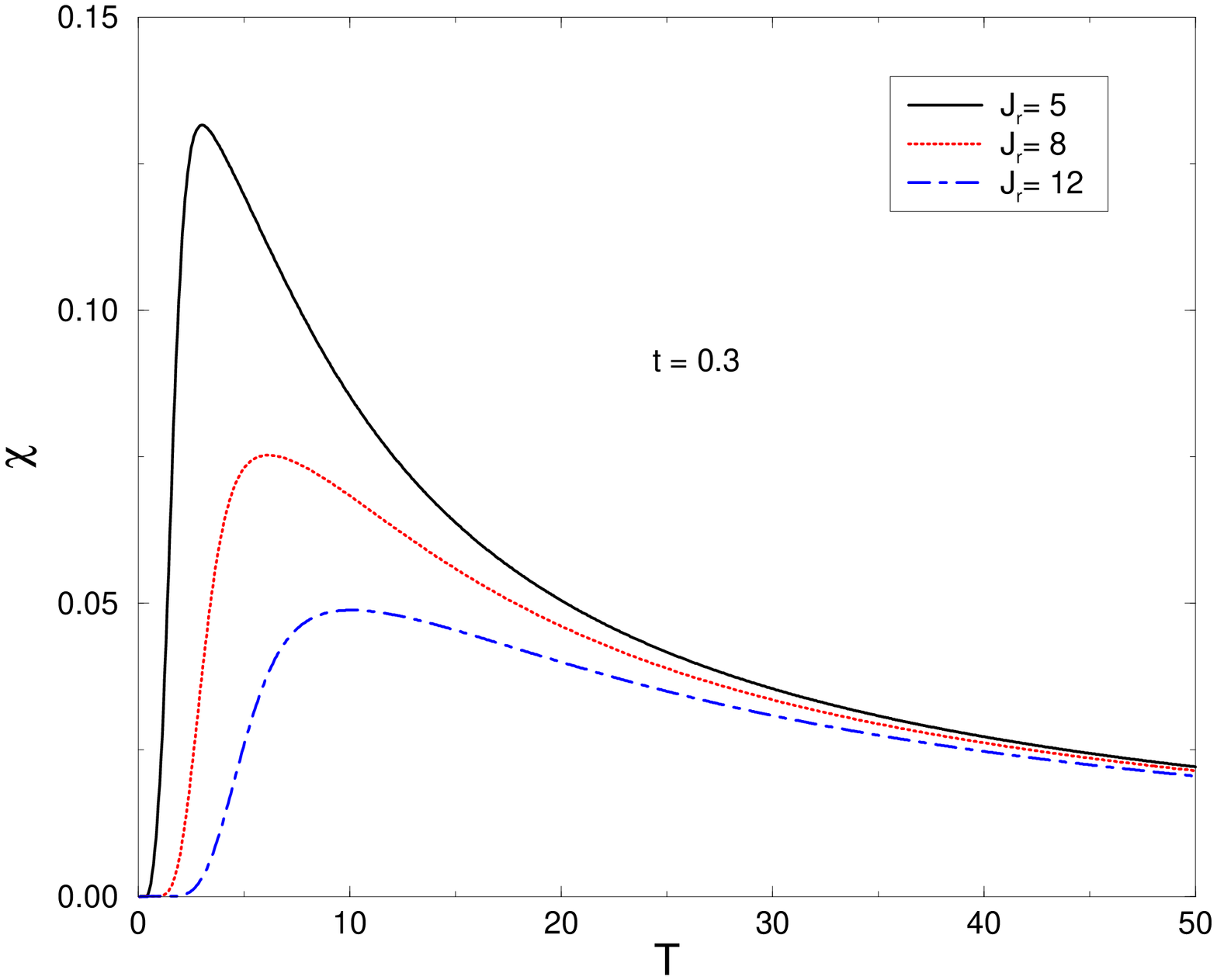}&            
\includegraphics[scale=0.3,angle=-90]{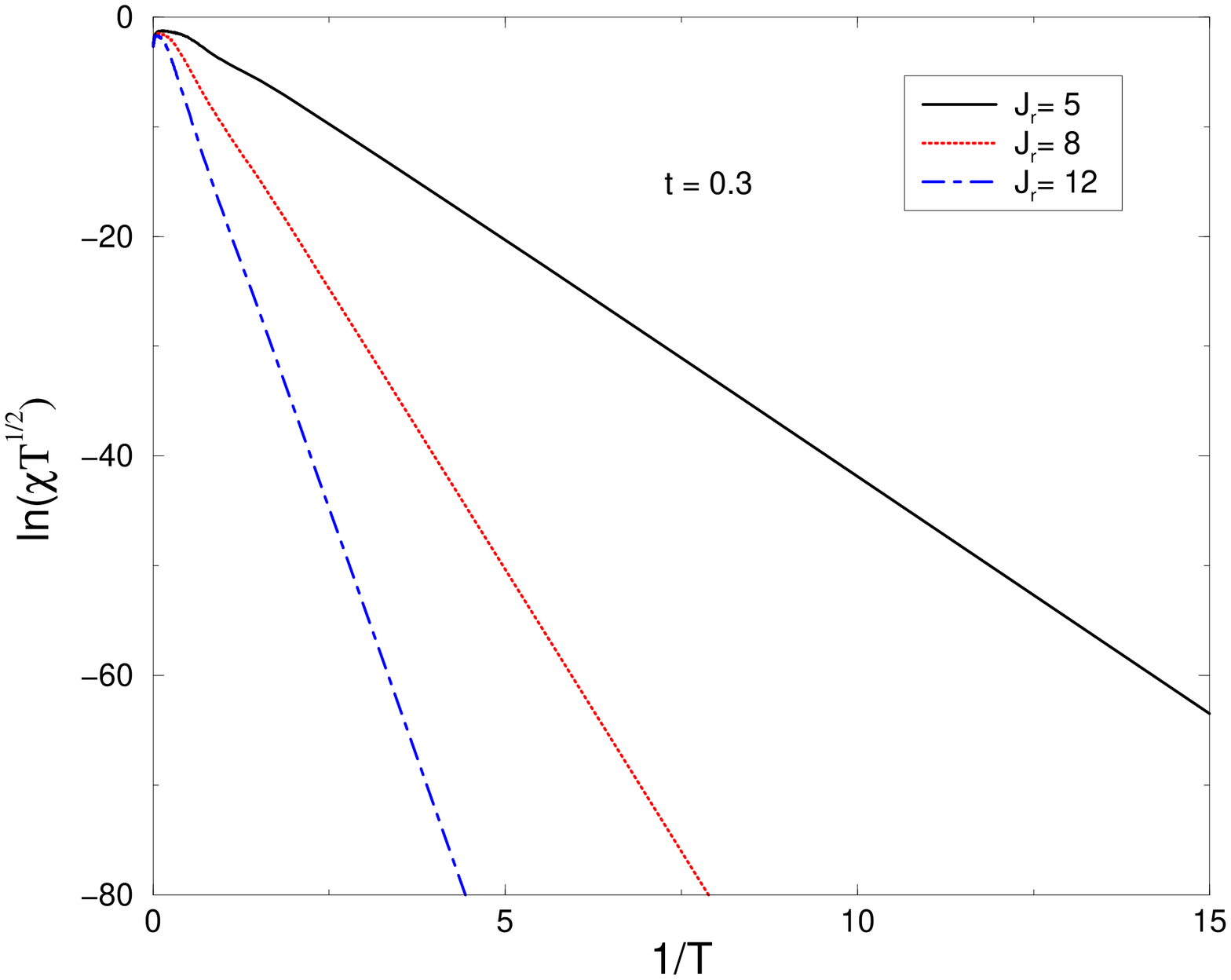} \\
          &               \\
\includegraphics[scale=0.3,angle=-90]{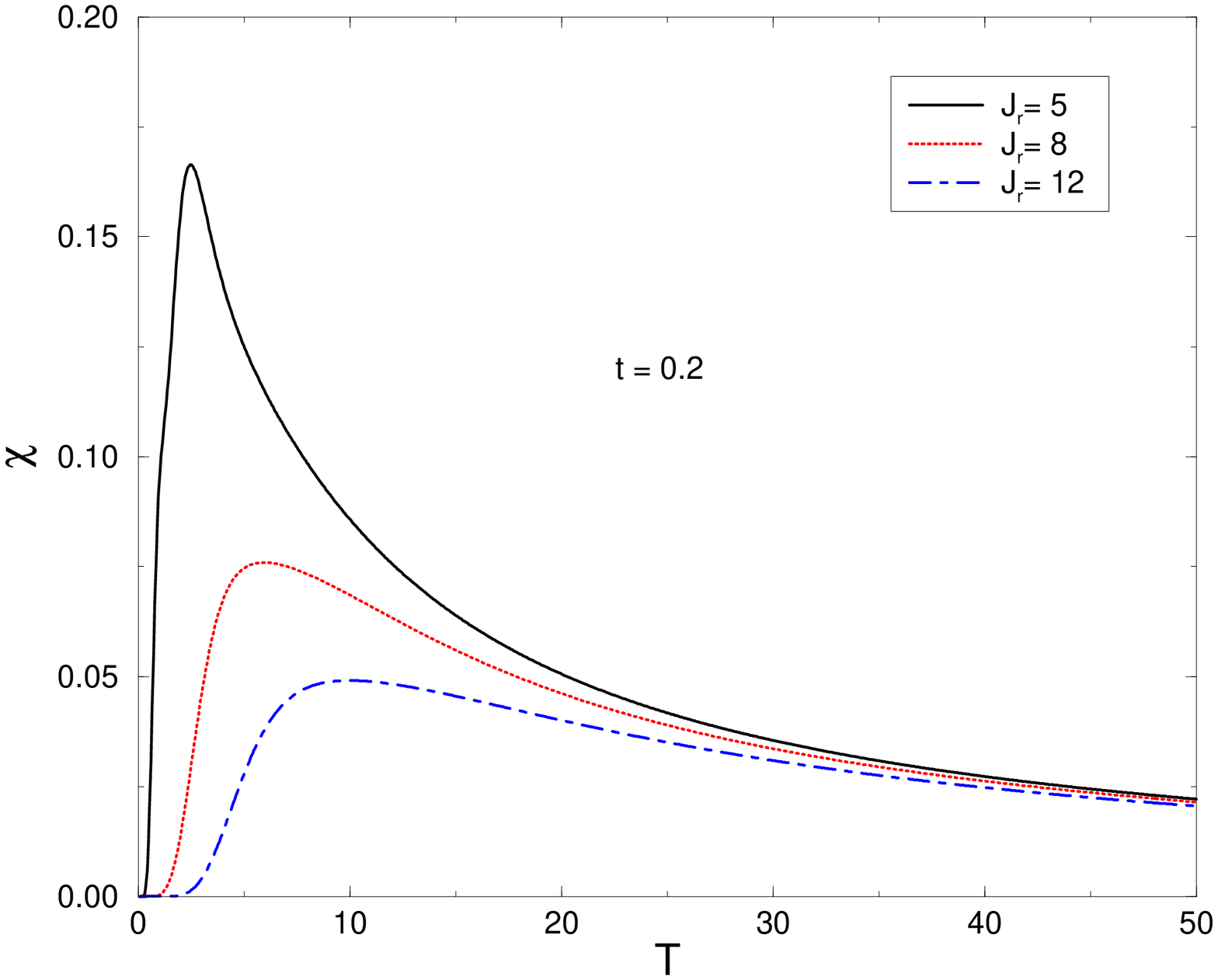}&            
\includegraphics[scale=0.3,angle=-90]{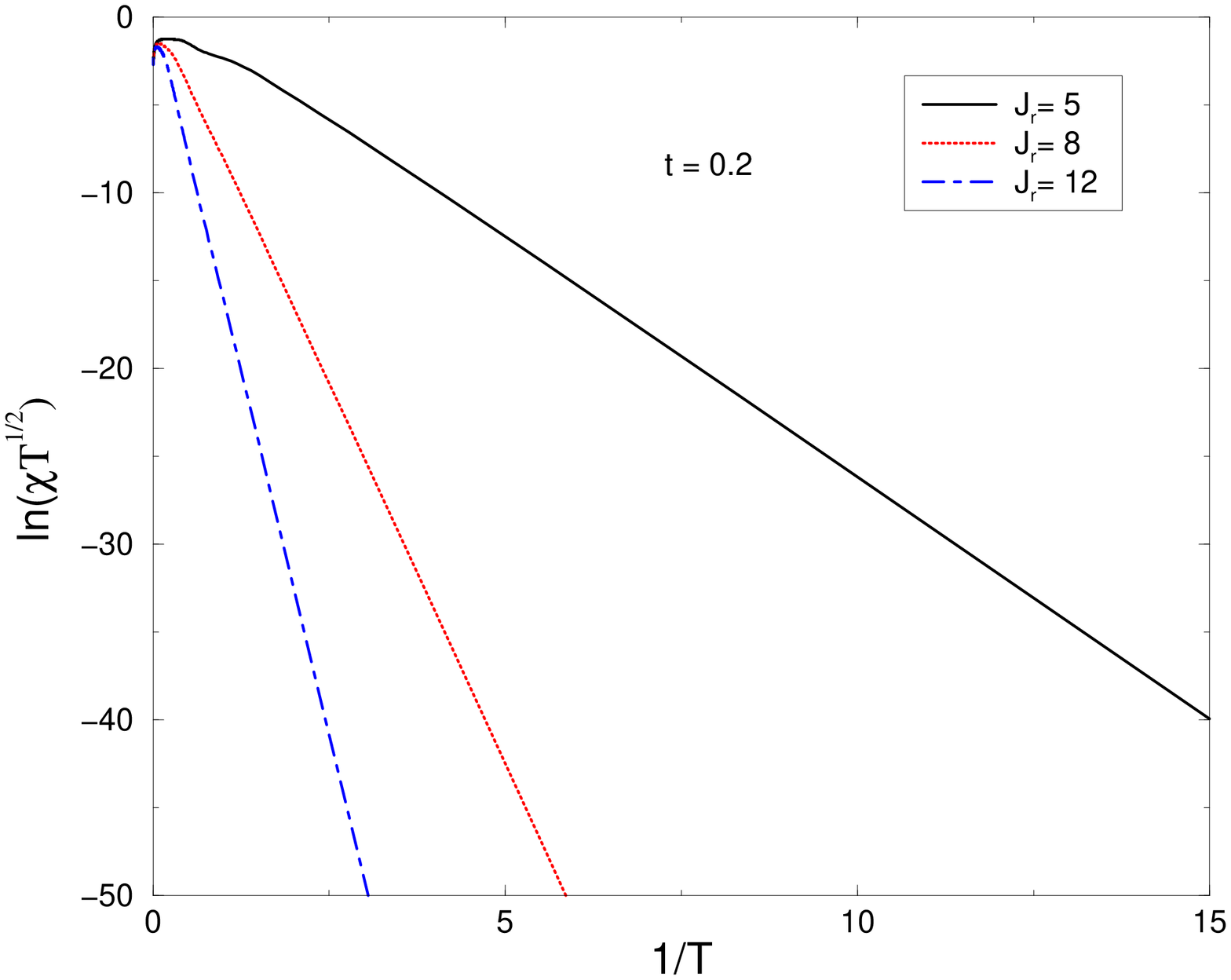} \\
\end{tabular}
\caption{ a) The magnetic susceptibility ($\chi$) as a function of the temperature ($T$) for different values of the coupling ($J_r$) and anisotropy ($t$).  
(b) A logarithmic plot of the  susceptibility ($\chi$)  as a function of the inverse of the  temperature ($1/T$), from which the spin gap ($\Delta^{\ast}$) can be obtained.} 
\end{center}
\end{figure}
The second point is that in all cases the susceptibility
presents an exponential decay for low temperature $T<<\Delta^{\ast}$
where $\Delta^{\ast}$ is the spin gap of the system  
\begin{equation}
\chi  \propto \frac{e^{-\frac{\Delta^{\ast}}{T}}}{\sqrt{T}}
\label{qui}
\end{equation}
which is in agreement with the result of Troyer {\it et. al.} \cite{r17}
for the Heisenberg spin ladder model.
By linearizing Eq. (\ref{qui})
a numerical value for the spin gap ($\Delta^{\ast}$) can be found.
When compared to the exact expression of the spin gap $\Delta$ obtained
from the Bethe Ansatz Equations (\ref{gap}) at $T=0$ an excellent agreement
is found as can be seen in Table $1$. The linearized curves are depicted
in Fig. 3b.
\begin{table}
\begin{center}
\begin{tabular}{|c|c|c|c|}\hline
 $ J_r$& t      &$\Delta$ &$ \Delta^{\ast} $ \\ \hline
     & 1.00 & 6.00 & 5.98 \\\cline{2-4}
     5& 0.30 & 4.37 & 4.42 \\ \cline{2-4}
      & 0.20 & 2.80  & 2.88 \\ \hline
      & 1.00 & 12.00 & 11.93 \\\cline{2-4}
    8 & 0.30 & 10.37 & 10.36 \\\cline{2-4}
      & 0.20 & 8.80 & 8.81 \\\hline
      & 1.00 & 20.00 & 19.88 \\ \cline{2-4}
   12 & 0.30 & 18.37 & 18.32 \\ \cline{2-4}
      & 0.20 & 16.80 & 16.76  \\ \hline
\end{tabular}
\caption{Spin gap $\Delta^{\ast}$ obtained from the linearization of Eq. (5)
and Fig. $3b$ compared to the analytical result $\Delta$ obtained from
the Bethe Ansatz Equations (4) for
different values of $J_r$ and $t$.}
\end{center}
\end{table}

We remark that the behavior of the magnetic susceptibility for the
critical value of the couplings (${J_r}^c$) is of the form

\begin{equation}
\chi \sim \frac{1}{\sqrt{T}}
\end{equation}
which indicates a typical quantum critical behavior. This had already been
predicted in \cite{r13,r14} for the isotropic case as illustrated
in Fig. $4$ for different values of $t$.
\begin{figure}[h]
\begin{center}
\includegraphics[scale=0.3,angle=-90]{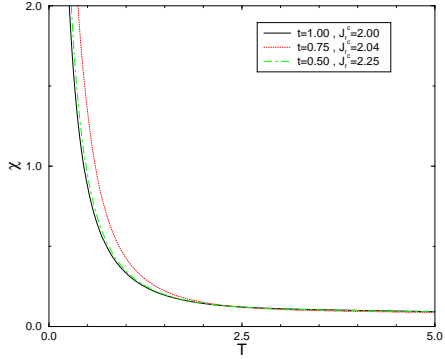}
\end{center}
\caption{Magnetic susceptibility ($\chi$) versus temperature ($T$) for
different values of the coupling $J_r$ in units of $J_l$.
These graphs indicate a
typical quantum critical behavior.}
\end{figure}
By setting our parameters accordingly we were able to obtain the
spin gap for some ladder compounds such as
$KCuCl_3$ \cite{r1,r2}, $Cu_2(C_5H_{12}N_2)_2Cl_4$ \cite{r3,r4,r5},
$(C_5H_{12}N)_2CuBr_4$ \cite{r6} which were previously studied by
different authors using different methods.
A comparison of our choice of parameters with the experimental data
$\Delta^{exp}$ \cite{r18} for the above mentioned compounds is depicted in
Table $2$.
\begin{table}
\begin{center}
\begin{tabular}{|c|c|c|c|c|c|}\hline
$compound $&$ J_r$& t  &$\Delta$ &$ \Delta^{\ast} $&$\Delta^{exp}$ \\ \hline
$KCuCl_3$ & 4.0 & 0.3254 & 32.00 & 36.56 & 31.10 \\\cline{1-6}
$Cu_2(C_5H_{12}N_2)_2Cl_4$ & 5.5 & 0.2344 & 10.80 & 13.41 & 10.80 \\\cline{1-6}
$(C_5H_{12}N)_2CuBr_4$ & 4.0 & 0.3100 & 9.85 & 9.90 & 9.50\\\hline 
\end{tabular}
\caption{The values of parameters for different compounds.
The experimental data were obtained from Ref. \cite{r18}. }
\end{center}
\end{table}
For the sake of clarity we present the results on three different compounds
separately and 
refer the reader to the quoted works
for more details on the experimental setup.
\subsection{$KCuCl_3$}

The magnetic properties of this system were investigated in \cite{r1,r2}
where a non-magnetic ground state 
and a spin gap in the excitation spectrum were reported. The
crystal structure of this compound is moniclinic and the double
chain feature of the system arises from two edge--sharing
chains of $CuCl_6$ octahedra (see \cite{r10} for more details on
the geometry of this compound).

The experimental data \cite{r1} and exact diagonalization results
are depicted in Fig. 5.
\vskip 0.5cm
\begin{figure}[h]
\begin{center}
\includegraphics[scale=0.3]{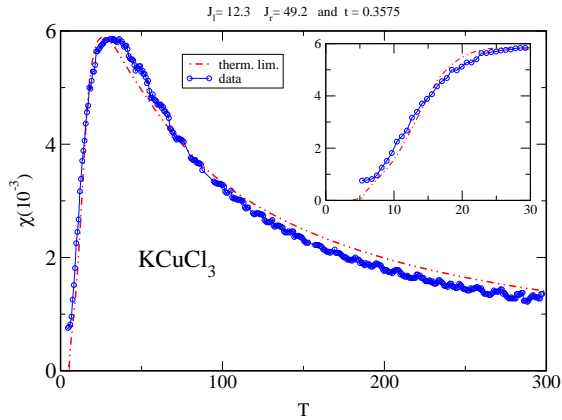}
\caption{ Magnetic susceptibility versus temperature. 
The full line is the result from extrapolation of
finite--size data obtained from exact diagonalization of chains
up to 12 spins. Circles are experimental data obtained 
by Nakamura and Okamoto \cite{r1}. The inset shows the
low--temperature regime.}
\end{center}
\end{figure}
The spin gap we used was estimated experimentally to be
$\sim 31.10 K$ \cite{r2}. The full line 
was obtained using the following exchange couplings: $J_l =12.3 K$,
$J_r=49.2 K$ and $t=0.3575$. We observe that the low--temperature
regime shows a better agreement between theory and experiment. The
deviation in the high--temperature regime might arise from the
existence, in the compound, of a ferromagnetic coupling which
takes place between the $3d$ orbitals of $Cu^{+2}$ ions mediated 
by the $Cl$ anions \cite{r20}, a coupling which is not present
in our model.

\subsection{ $Cu_2(C_5H_{12}N_2)_2Cl_4$}

This compound have been studied by Chaboussant et al \cite{r4,r5,r21 }.
The ladder structure arises from the stacking of $Cu$--$Cu$ binuclear
units (each $Cu$ is to be seen as belonging to one leg). For
magnetic fields below $h = 7.5 T$ the one--dimensional ground state is a
valence-bond singlet structure on each rung and the system is gapped.
Above this lower critical field, the first excited state (a triplet)
becomes the ground state and the system has a non--zero magnetization.
There are basically three different exchange paths between $Cu$ ions:
the intramolecular exchange
constant $J_r$ between $Cu$-$Cu$ ions takes place through two $Cl$
ions while the intermolecular $J_l$ involves
in--plane unpaired electron densities of $Cu$ ions of different
molecules. The third intermolecular exchange introduces a frustration
via next--nearest neighbor interaction and occurs through hydrogen
bonds between the $Cl$ and $N$ ions.
This interaction is however significantly smaller
than $J_r$ and is taken altogether equal to zero in our model.
The low--temperature gap has been determined through susceptibility
measurements and NMR techniques to be $\Delta = 10.8
\pm 0.6 $K with $J_l= 2.4$K and $J_r= 13.2$K \cite{r4}. The same
gap can be obtained in our model with the choices
$J_l=2.4 K$, $J_r=13.2 K$ and $t=0.2344$. The agreement between
the low--temperature experimental data and our results is excellent, as
can be seen in Fig. 6, and remains
good for the whole temperature range.
\vskip 0.5cm
\begin{figure}[h]
\begin{center}
\includegraphics[scale=0.3]{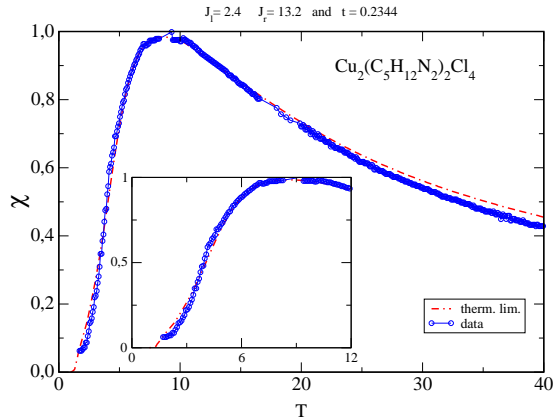}
\caption{ Normalized magnetic susceptibility versus temperature. 
The full line is the result of exact diagonalization followed by
extrapolation of finite--size data. Again we used chains of up to
12 spins. The circles are the
experimental data obtained by Chaboussant et al \cite{r4}. The inset
depicts the low--temperature regime.}
\end{center}
\end{figure} 
\subsection{ $(C_5H_{12}N)_2CuBr_4$ }

As in the previous compound, here we also have a binuclear $Cu$--$Cu$ stacked
structure. The intramolecular exchange coupling is mediated through an
orbital overlap of $Br$ ions adjacent to the $Cu$ sites. The intermolecular
exchange is also mediated by $Br$ ions a bit further apart ($8.597 \AA$
in comparison to intramolecular distances of $6.934\AA$). A diagonal
frustration exchange should be present but it is expected to be
weak \cite{r6}. 
The exchange couplings used were $J_l= 4.0 K$, $J_r = 16.0 K$ and $t=0.31K$.
Again the fit is better for low temperatures, with an energy gap of
$9.8$ K ($9.5$ K experimentally \cite{r6}) and an exponential drop in
the magnetic susceptibility. The data is depicted in Fig. 7.
\begin{figure}[h]
\begin{center}
\includegraphics[scale=0.3]{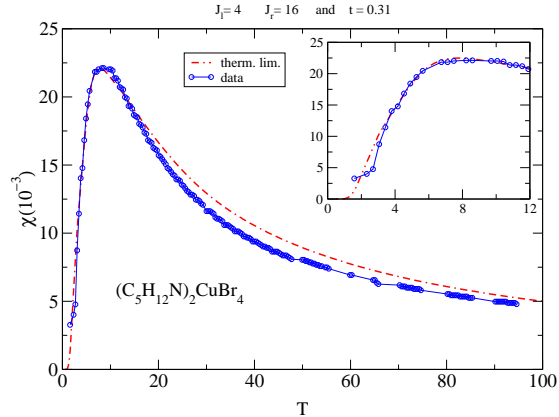}
\caption{ Magnetic susceptibility versus temperature. Circles
represent experimental data obtained by Watson {\it et. al.} \cite{r6}.
The solid line is the result obtained via exact diagonalization
as explained in the previous captions. The inset depicts the low--temperature
regime.}
\end{center}
\end{figure}
%

\section{Summary}
We presented a numerical analysis of thermodynamic properties of an 
integrable anisotropic spin ladder model which has, besides the rung
and leg couplings, an extra free anisotropy parameter.
From the Bethe Ansatz equations as well as the exact expression for
the energy spectrum \cite{r12} we showed that the model exhibits an
anisotropy-- dependent spin gap. The critical line and phase
diagram were obtained.
The magnetic susceptibility was investigated and good agreement was
found between the gap obtained from the Bethe ansatz equations and the gap
obtained from the magnetic susceptibility. A connection with some
strong coupling compounds was discussed and our results reproduce
well experimental data for these compounds in strong coupling limit.
The anisotropy parameter in our model has no obvious physical
interpretation. However, cyclic four-spin exchange terms are known to
affect the excitation spectrum of spin ladders and the importance of
this type of interaction for appropriately describing the properties
of strong--coupling ladders has also been established \cite{nunner}.
As shown in this work the anisotropy parameter allows us to tune
our model to different physical systems of interest, thus
presenting a unified scenario for strong--coupling ladders. 

This work has been financially supported by the brazilian federal
agency CNPq (APT, AF and ALM) and the state agency FAPESP (ALM).


\end{document}